\documentclass{article}
\usepackage[psamsfonts]{amssymb}
\usepackage{amsmath}
\usepackage{cite}
\usepackage{bm}
\usepackage{graphicx}
\usepackage{yhmath}
\usepackage{bbm}

\def\d{\mathrm{d}}

\def\id{\mathbf{1}}

\def\ir{\mathrm{i}}

\begin{document}
\begin{titlepage}
\begin{center}
{\large \textbf{
Eigenvalue problem for radial potentials
\vskip 0.3cm
 in space with SU(2) fuzziness}}

\vspace{2\baselineskip}

Marjan-S. Mirahmadi
~~~~~~and~~~~~~
Amir~H.~Fatollahi~\footnote {fath@alzahra.ac.ir}
\\
\vskip 10 mm
\textit{ Department of Physics, Alzahra University, Tehran
1993891167, Iran }
\end{center}
\vspace{\baselineskip}
\begin{abstract}
\noindent The eigenvalue problem for radial potentials is considered in a space whose
spatial coordinates satisfy the SU(2) Lie algebra. As the consequence,
the space has a lattice nature and the maximum value of momentum is bounded 
from above. The model shows interesting features due to the bound, namely,
a repulsive potential can develop bound-states, or an attractive
region may be forbidden for particles to propagate with higher energies.
The exact radial eigen-functions in momentum space are given by means of
the \textit{associated Chebyshev functions}. For the radial stepwise potentials
the exact energy condition and the eigen-functions are presented.
For a general radial potential it is shown that the discrete energy spectrum
can be obtained in desired accuracy by means of given forms of
continued fractions.
\end{abstract}

\vspace{2\baselineskip}

\textbf{PACS numbers:} 02.40.Gh, 03.65.-w, 03.65.Ge

\textbf{Keywords:} Noncommutative geometry; Quantum mechanics; Bound-state wave equations

\end{titlepage}

\section{Introduction}
The noncommutative spaces have been the subject of a great number of
studies in recent years \cite{doplicher,madore1}. The natural appearance of these
spaces in some areas of physics, for example in the string theory,
is a part of the motivation. In particular, the canonical relation
\begin{equation}\label{1}
[\hat x_a,\hat x_b]=\mathrm{i} \,\theta_{a\, b}\,\id,
\end{equation}
is shown to describe the algebra between the coordinates
of the longitudinal directions of $D$-branes
in presence of a constant $B$-field background
\cite{9908142,99-2,99-3,99-4,reviewnc}.

The natural extension of the above algebra is to take
the commutators of the coordinates non-constant.
Examples of this kind are,
the noncommutative cylinder and the $q$-deformed plane \cite{chai},
the $\kappa$-Poincar\'{e} algebra \cite{majid,ruegg,amelino,kappa},
and linear noncommutativity of the Lie algebra type \cite{wess,sasak}.
In the latter the dimensionless spatial position operators satisfy:
\begin{equation}\label{2}
[\hat{x}_a,\hat{x}_b]= f^c{}_{a\, b}\,\hat{x}_c,
\end{equation}
in which $f^c{}_{a\,b}$'s are the structure constants of a Lie algebra,
for example the algebra by SO(3) or SU(2) groups.
A special case, the so-called fuzzy sphere,
is when an irreducible representation of the position operators is
taken, by which the Casimir of the algebra is constant, hence the name
sphere \cite{madore,presnaj}.

The other possibility is to not restrict the representation
to an irreducible one, but all of the irreducible representations
would be taken \cite{0612013,fakE1,fakE2,spga}; see also \cite{jablam}.
In particular, the regular representation of the group would be considered,
and as the consequence, the model is built on the whole space,
not on a sub-space, as the case with fuzzy sphere.

In \cite{0612013,fakE1,fakE2,spga} basic
ingredients for calculus on a linear Lie type fuzzy space and
the field theoretic aspects on such a space were studied in details.
The most remarkabale features of the field theories on such
a space happen to be: 1) They are free from any ultraviolet divergences
if the group is compact; 2) The momentum conservation is modified,
in the sense that the vector addition is replaced by a non-Abelian
operation \cite{pal,0612013};
3) In the transition amplitudes only the so-called planar graphs contribute.

The classical motion on noncommutative space
has attracted interests as well \cite{miao,silva}.
In particular, the central force problems on space-times
with canonical and linear noncommutativity and their
observational consequences have been the subject of different
research works \cite{levia,zhang,kapoor,dennis,romero,mirza}.
In \cite{kfs} the classical mechanics defined on a space with
SU(2) algebra was studied. In particular, the Poisson structure
induced by noncommutativity of SU(2) type was investigated, for
either the Cartesian or Euler parametrization of SU(2) group.
In \cite{andalib} it was shown that on a SU(2) type space
it is only the Kepler potential, as a single-term power-law one,
for which all of nearly circular orbits are closed.
Further, it was proved for the Kepler potential all of
bounded orbits, no matter how far from circle,
is closed \cite{andalib}.

The commutation relations of the position and momentum
operators on a space with SU(2) algebra was studied in \cite{fsk}.
The thermodynamical aspects of these models were explored in \cite{shin,fsmjmp}.

The purpose of the present work is to continue the study the
quantum mechanics on space with SU(2) algebra. In particular,
the eigenvalue problem for radial potentials is considered in a space whose
spatial coordinates satisfy the SU(2) Lie algebra. The interesting feature
by the SU(2) algebra is, the space happens to have a lattice nature,
however a rotationally symmetric one. Also, due to the lattice structure,
the maximum value of momentum is bound from above.
The bound on momentum would appear as the basis for
surprising features for the model. In particular, on such a space
a repulsive potential can develop bound-states, or an attractive
region may be forbidden for particles to propagate with higher energies.
As definition of the position eigenstates of the form
$|x_1\rangle |x_2\rangle|x_3\rangle$ is not possible
due to the algebra (\ref{2}), the momentum space is commutative, and hence
all the necessary ingredients for the model can be defined in this space.

The scheme of the rest of this paper is as following. In Sec.~2,
the basic notions to formulate quantum theory
on a space with Lie type noncommutativity are presented.
Also in this section the construction is specialized for the case of the
SU(2) group. In Sec.~3 the exact radial eigen-functions are constructed
by means of associated Chebyshev functions. In Sec.~4
the radial stepwise potentials are considered, and the exact
expression for the energy quantization condition as well as
the eigen-functions are presented. In Sec.~5 the case
with a general radial potential is discussed. In particular
 it is shown that the discrete energy spectrum
can be obtained in desired accuracy by means of given forms of
continued fractions.

\section{Basic notions}
Consider a Lie group G. Denote the members of a basis for the
left-invariant vector fields corresponding to this group by
$\hat{x}_a$'s. These fields (which are sections of the tangent
bundle TG) satisfy (\ref{2}), with the structure constants of
the Lie algebra corresponding to G. The coordinates ${k}^a$
are defined such that
\begin{equation}\label{3}
U({\mathbf{k}}):=\exp({k}^a\,\hat{x}_a)\,U(\mathbf{0}),
\end{equation}
where $U({\mathbf{k}})$ is the group element corresponding to
the coordinates ${\mathbf{k}}$, $U(\mathbf{0})$ is the
identity, and $\exp(\hat{x})$ is the flux corresponding to the
vector field $\hat{x}$. The Hilbert space to be considered is the
space of functions defined on G, which are square integrable with
respect to the Haar measure of the group. The action of the
functions of the group and the vector fields defined on the group,
on the functions of the group are defined through multiplication
and Lie derivation, respectively.
The commutators of the operator forms of the coordinate functions and
the left invariant vector fields are (\ref{2}) and
\begin{align}\label{4}
[\hat{k}^a,\hat{k}^b]&=0, \\\label{5}
[\hat{x}_a,\hat{k}^b]&=\hat{x}_a{}^b,
\end{align}
where $\hat{x}_a{}^b$'s are functions of G. These satisfy
\begin{equation}\label{6}
\hat{x}_a{}^b(\hat{\mathbf{k}}=\mathbf{0})=\delta_a^b.
\end{equation}

Next consider the right-invariant vector fields
$\hat{x}_a^{\mathrm{R}}$, so that they coincide with their
left-invariant analogues at the identity of the group \cite{fsk}:
\begin{equation}\label{7}
\hat{x}_a^{\mathrm{R}}(\hat{\mathbf{k}}=\mathbf{0})=
\hat{x}_a(\hat{\mathbf{k}}=\mathbf{0}).
\end{equation}
These field satisfy the commutation relations
\begin{align}\label{8}
[\hat{x}_a^{\mathrm{R}},\hat{x}_b^{\mathrm{R}}]&=
-f^c{}_{a\,b}\,\hat{x}_c^{\mathrm{R}},\\
\label{9}
[\hat{x}_a^{\mathrm{R}},\hat{x}_b]&=0.
\end{align}
Using these, one defines the new vector field $\hat{J}_a$ through
\begin{equation}\label{10}
\hat{J}_a:=\hat{x}_a-\hat{x}_a^{\mathrm{R}}.
\end{equation}
These are the generators of the adjoint action, and satisfy the
commutation relations \cite{fsk}
\begin{align}\label{11}
[\hat{J}_a,\hat{J}_b]&=f^c{}_{a\,b}\,\hat{J}_c,\\
\label{12} [\hat{J}_a,\hat{x}_b]&=f^c{}_{a\,b}\,\hat{x}_c,\\
\label{13} [\hat{J}_a,\hat{x}_b^{\mathrm{R}}]&=
f^c{}_{a\,b}\,\hat{x}_c^{\mathrm{R}},\\
\label{14} [\hat{k}^c\,\hat{J}_a]&=f^c{}_{a\,b}\,\hat{k}^b.
\end{align}
For the group SU(2), taking $\hat{k}^a$'s and $\hat{x}_a$'s as
momenta and spatial coordinates respectively, $\hat{J}_a$'s are
the natural candidates for the orbital angular momenta, as suggested by
the algebra they satisfy.

Using the dimensionless operators introduced in the above, one can easily
construct the corresponding dimensionful ones, simply by
multiplication of these operators by suitable factors to make them
Hermitian with proper dimension:
\begin{align}\label{15}
p^a&:=(\hbar/\ell)\,\hat{k}^a,\\
\label{16} x_a&:=\ir\,\ell\,\hat{x}_a,\\
\label{17} x_a^{\mathrm{R}}&:=\ir\,\ell\,\hat{x}_a^{\mathrm{R}},\\
\label{18} x_a{}^b(\mathbf{p})&:=
\hat{x}_a{}^b[(\ell/\hbar)\,\mathbf{p}],\\
\label{19} J_a&:=\ir\,\hbar\,\hat{J}_a,
\end{align}
where $\ell$ is a constant of dimension length. One then arrives
at the following commutation relations \cite{fsk}
\begin{align}\label{20}
[p^a,p^b]&=0,\\
\label{21} [x_a,p^b]&=\ir\,\hbar\,x_a{}^b,\\
\label{22} [x_a,x_b]&=\ir\,\ell\,f^c{}_{a\,b}\,x_c,\\
\label{23} [J_a,x_b]&=\ir\,\hbar\,f^c{}_{a\,b}\,x_c,\\
\label{24} [p^c,J_a]&=\ir\,\hbar\,f^c{}_{a\,b}\,p^b,\\
\label{25} [J_a,J_b]&=\ir\,\hbar\,f^c{}_{a\,b}\,J_c,
\end{align}
It is seen that in the limit $\ell\to 0$ the ordinary commutation
relations are retrieved.

\subsection{SU(2) setup and the Euler parameters}

For the group SU(2), the commutation relations (\ref{20}),
(\ref{21}), and (\ref{22}) make in fact the algebra of a
rigid rotator, in which the angular momentum and the rotation
vector have been replaced by $\mathbf{x}$ and $\mathbf{p}$,
respectively, that is, the roles of position and momenta have been
interchanged. As the consequence, the position operators do not
have simultaneous eigenstates and the space has a
lattice structure. As the momentum space is commutative with well-defined
eigenstates, we switch to this space. As usual it is convenient to use the
Euler parametrization of SU(2), defined through
\begin{equation}\label{26}
\exp(\phi\,T_3)\,\exp(\theta\,T_2)\,\exp(\psi\,T_3):=
\exp({k}^a\,T_a),
\end{equation}
where $T_a$'a are the generators of SU(2) satisfying the relations
\begin{align}\label{27}
[T_a,T_b]&=\epsilon^c{}_{a\,b}\,T_c,\\\label{28}
[T_a,T_b]_+&=-\frac{1}{2}\,\delta_{a\,b},
\end{align}
for which the second is valid for the defining representation of SU(2).
It can be seen that the range of the Euler parameters so that each
point of the group is covered one and only one time is \cite{fsk}
\begin{align}\label{29}
0&\leq\phi+\psi\leq 2\,\pi,\nonumber\\
-2\,\pi&\leq\phi-\psi\leq 2\,\pi,\nonumber\\
0&\leq\theta\leq 2\,\pi.
\end{align}
One also has \cite{goldstein}
\begin{equation}\label{30}
\cos \frac{k}{2} = \cos \frac{\theta}{2}\,
\cos\frac{\phi+\psi}{2},
\end{equation}
where $k:=\sqrt{\delta_{a\,b}\,{k}^a\,{k}^b}$.
In the Euler momentum basis the inner-product of wave-functions
is defined using the so-called Haar measure $\d\mu$, given by:
\begin{equation}\label{31}
\d\mu=c\,|\sin\theta|\,\d\phi\,\d\theta\,\d\psi,
\end{equation}
in which $c$ is a constant, and is fixed once the normalization
prescription is fixed.

The operators $\hat{\mathbf{x}}$ and $\hat{\mathbf{J}}$ in the momentum basis
with proper dimension are given in \cite{fsk}:
\begin{align}\label{32}
x_1\to&\;\ir\,\ell\,\left(-\frac{\cos\psi}{\sin\theta}\,
\frac{\partial}{\partial\phi}+\sin\psi\,\frac{\partial}{\partial\theta}+
\frac{\cos\psi\,\cos\theta}{\sin\theta}\,\frac{\partial}{\partial\psi}\right),\\
\label{33}
x_2\to&\;\ir\,\ell\,\left(\frac{\sin\psi}{\sin\theta}\,
\frac{\partial}{\partial\phi}+\cos\psi\,\frac{\partial}{\partial\theta}-
\frac{\sin\psi\,\cos\theta}{\sin\theta}\,\frac{\partial}{\partial\psi}\right),\\
\label{34}
x_3\to&\;\ir\,\ell\,\frac{\partial}{\partial\psi},
\end{align}
\begin{align}
\label{35}
J_1\to&\;\ir\,\hbar\,\left[\frac{\cos\phi\,\cos\theta-\cos\psi}{\sin\theta}\,
\frac{\partial}{\partial\phi}+(\sin\phi+\sin\psi)\,
\frac{\partial}{\partial\theta}\right.\nonumber\\
&\left.+\frac{-\cos\phi+\cos\psi\,\cos\theta}{\sin\theta}\,\frac{\partial}{\partial\psi}\right],\\
\label{36}
J_2\to&\;\ir\,\hbar\,\left[\frac{\sin\phi\,\cos\theta+\sin\psi}{\sin\theta}\,
\frac{\partial}{\partial\phi}+(-\cos\phi+\cos\psi)\,
\frac{\partial}{\partial\theta}\right.\nonumber\\
&\left.+\frac{-\sin\phi-\sin\psi\,\cos\theta}{\sin\theta}\,\frac{\partial}{\partial\psi}\right],\\
\label{37}
J_3\to&\;\ir\,\hbar\,\left(-\frac{\partial}{\partial\phi}+
\frac{\partial}{\partial\psi}\right).
\end{align}
It can be shown that the above operators are Hermitian with respect
to the inner-product defined by the Haar measure (\ref{31}).

Introducing the new parameters:
\begin{align}\label{38}
\chi&:=\frac{\phi-\psi}{2},~~~~~~~~~ \xi:=\frac{\phi+\psi}{2},\nonumber\\
v&:=\cos\frac{\theta}{2}\,\cos\xi,~~~~ \tau:=(1-v^2)^{-1/2}\,\cos\frac{\theta}{2}\,\sin\xi,
\end{align}
one arrives at ($J_\pm=J_1\pm\ir\,J_2$)
\begin{align}\label{39}
J_\pm&=\ir\,\hbar\,\exp(\pm\,\ir\,\chi)\,\left(-\sqrt{1-\tau^2}\,
\frac{\partial}{\partial\tau}\pm\,\ir\,\frac{\tau}{\sqrt{1-\tau^2}}\,
\frac{\partial}{\partial\chi}\right),
\\ \label{40}
J_3&=-\ir\,\hbar\,\frac{\partial}{\partial\chi},
\end{align}
resulting in
\begin{align}\label{41}
\mathbf{J}\cdot\mathbf{J}&=-\hbar^2\,\left[(1-\tau^2)\,
\frac{\partial^2}{\partial\tau^2}-2\,\tau\,
\frac{\partial}{\partial\tau}+\frac{1}{1-\tau^2}\,
\frac{\partial^2}{\partial\chi^2}\right],
\end{align}
and subsequently \cite{fsk}:
\begin{align}\label{42}
\mathbf{x}\cdot\mathbf{x}=
-\frac{\ell^2}{4}\left[-\frac{\hbar^{-2}}{1-v^2}\,
\mathbf{J}\cdot\mathbf{J}+
(1-v^2)\frac{\partial^2}{\partial
v^2}-3v\,\frac{\partial}{\partial v}\right].
\end{align}

Using (\ref{40}) and (\ref{41}), it is seen that the
angular momentum eigenfunctions ($\mathcal{Y}_l^m$'s) satisfying
\begin{align}\label{43}
J_3\,\mathcal{Y}_l^m&=m\,\hbar\,\mathcal{Y}_l^m, \\
\label{44}
\mathbf{J}\cdot\mathbf{J}\,\mathcal{Y}_l^m&=l\,(l+1)\,\hbar^2\,
\mathcal{Y}_l^m,
\end{align}
are products of an arbitrary function $f(v)$, and $Y_l^m$'s
(the usual spherical harmonics) with the cosine of the colatitude
equal to $\tau$ and the longitude equal to $\chi$, that is
\begin{equation}\label{45}
\mathcal{Y}_l^m=f(v)\,Y_l^m(\cos^{-1}\tau,\chi).
\end{equation}

Hereafter we consider SU(2)-invariant systems, that is they
are rotationally invariant and the Hamiltonian $H$ and $J_a$'s commute.
As $J_a$'s generate rotations of both
$\mathbf{x}$ and $\mathbf{k}$, for a SU(2)-invariant system
$H$ is a function of only $\mathbf{p}\cdot\mathbf{p}$ and
$\mathbf{x}\cdot\mathbf{x}$, namely:
\begin{equation}\label{46}
H=K(\sqrt{\mathbf{p}\cdot\mathbf{p}})+V(\sqrt{\mathbf{x}\cdot\mathbf{x}})
\end{equation}
in which $K$ and $V$ are representing the kinetic and the potential terms,
respectively. An example for $K$ is \cite{fakE1,fakE2,fsk,kfs}
\begin{align}\label{47}
K&=\frac{4\,\hbar^2}{M\,\ell^2}\,
\left(1-\cos\frac{\ell\,p}{2\,\hbar}\right),\nonumber\\
&=\frac{4\,\hbar^2}{M\,\ell^2}\,(1-v).
\end{align}
By the above choice, originated from the characteristics of spin-half irreducible
representations of the group, the kinetic term happens to be
monotonic with respect to $k$ (for $0<k<2\,\pi$) \cite{fakE1,fakE2,fsk,kfs}.
In the commutative limit $\ell\to 0$ this kinetic term is reduced to
the commutative case $\mathbf{p}\cdot\mathbf{p}/(2\,M)$.

For such a SU(2)-invariant system, $H$,
$\mathbf{J}\cdot\mathbf{J}$ and one of the components of
$\mathbf{J}$ (say $J_3$) can be taken to have common
eigen-functions. Now, the aim is to exploit the SU(2)-symmetry of such a
Hamiltonian to write down an eigenvalue equation for the
Hamiltonian so that that equation contains only one variable, out of the
the three variables corresponding to the momentum. It is in fact an easy task
by the expressions obtained so far. By the given form
of $\mathbf{x}\cdot\mathbf{x}$ by (\ref{42}), and the relation (\ref{44}),
one finds \cite{fsk}
\begin{equation}\label{48}
\mathbf{x}\cdot\mathbf{x}~\mathcal{Y}_l^m=Y_l^m\,\frac{\ell^2}{4}\left[
- (1-v^2) \frac{\d^2}{\d
v^2}+3v\, \frac{\d}{\d v}+\frac{l\,(l+1)}{1-v^2}\right] \,f(v),
\end{equation}
by which the radial part of the Schrodinger equation in momentum
space takes the form
\begin{align}\label{49}
\frac{4\,\hbar^2}{M\,\ell^2}\,(1-v)\,\psi_{Elm}(v)+
V(\sqrt{\mathbf{x}\cdot\mathbf{x}})
 \psi_{Elm}(v) =E\,\psi_{Elm}(v).
\end{align}


\section{Radial eigenfunctions}
As seen in Sec.~2, for the systems with rotational invariance, as in case on ordinary
space, the eigenvalue problem is reduced to a one-dimensional one. However,
it is reminded that in the present case the one-dimensional problem,
in contrary to ordinary space is not the length of the position vector
$r=\sqrt{\mathbf{x}\cdot\mathbf{x}}$, but it is the length of
momentum vector, $p=\sqrt{\mathbf{p}\cdot\mathbf{p}}$,
or alternatively $v=\cos (\ell\, p/2\hbar)$ with $-1\leq v\leq 1$. So it is natural
to define the basis $|v\rangle_l$ as the eigenvector of operator $\hat{v}$
acting on subspace with orbital angular momentum $l$:
\begin{equation}\label{50}
\hat{v}\, |v\rangle_l = v\, |v\rangle_l ,
\end{equation}
for which using the Haar measure (31) and the variables (\ref{38}), we have
\begin{equation}\label{51}
{}_l\langle v |v'\rangle_l = \frac{\delta{(v-v')}}{\sqrt{1-v^2}}.
\end{equation}
In the present section the aim is to find the eigen-functions of the
operator $\mathbf{x}\cdot\mathbf{x}$, for which
in the $v$-space we earlier found:
\begin{equation}\label{52}
\mathbf{x}\cdot\mathbf{x}=-\frac{\ell^2}{4}\left[
 (1-v^2) \frac{\d^2}{\d v^2}
-3v \frac{\d}{\d v}-\frac{l(l+1)}{1-v^2}\right]
\end{equation}
Fortunately the operator in the bracket for $l=0$ is known,
with the \textit{Chebyshev polynomials} of Type~II as
eigen-functions, satisfying:
\begin{equation}\label{53}
 (1-v^2)\,U''_n(v)-3v\, U'_n(v)=-n(n+2)\, U_n(v).
\end{equation}
Constructing the eigenfunctions for cases with $l\neq 0$ is
rather straightforward, just like the method by which
the associated Legendre polynomials are constructed \cite{arfken}.
In general, we will find for the \textit{associated Chebyshev functions} the following
\begin{equation}\label{54}
(1-v^2)\,{U_n^l}''(v)-3v\, {U_n^l}'(v)-\frac{l(l+1)}{1-v^2}U_n^l(v)=-n(n+2)\, U_n^l(v)
\end{equation}
in which $n=l,l+1,\cdots$, and
\begin{equation}\label{55}
U_n^l(v)= \sqrt{\frac{2}{\pi}\frac{(n-l)!(n+1)}{(n+l+1)!}} (1-v^2)^{l/2} \frac{\d^l}{\d v^l}U_n(v),
\end{equation}
in which the pre-factor is set in the way that the eigen-functions
are normalized, satisfying
\begin{equation}\label{56}
\int_{-1}^1 \sqrt{1-v^2}~ U_n^l(v)~ U_{n'}^l(v) = \delta_{nn'}.
\end{equation}
It is reminded that the original $U_n(v)$ is not normalized to one,
and in fact $U_n^0=\sqrt{\frac{2}{\pi}}\,U_n$.
Also, as $U_n$ is a polynomial of degree $n$,
by construction $U_n^l\equiv 0$, for $n< l$.
As the associated Chebyshev functions are rather less available,
in the Appendix~A explicit expressions for them together with the plots are presented.
Readily, by the basis constructed by $U_n^l$'s, a representation
of $\delta$-function in $v$-space is given
\begin{equation}\label{57}
\sum_{n=l}^\infty U_n^l(v) U_n^l(v')= \frac{\delta(v-v')}{\sqrt{1-v^2}}.
\end{equation}
By these all, the eigenvalues of the operator
$\mathbf{x}\cdot\mathbf{x}$ happen to be $\frac{n}{2}(\frac{n}{2}+1)\ell^2$,
for $l=0,1,\cdots,n$, leading to the degeneracy $2(\frac{n}{2})+1$. Reminding
that the coordinates $\hat{x}_a$'s satisfy the algebra (\ref{22}) for SU(2), this result
is the one to be expected. So every wave-function with orbital angular momentum $l$
can be expanded in terms
of $U_n^l$'s, namely
\begin{equation}\label{58}
\psi_l(v)=\sum_{n=l}^\infty a_n\, U_n^l(v).
\end{equation}
By the above expansion, the probability that the particle would be found at
the radial site $r_n=\sqrt{\frac{n}{2}(\frac{n}{2}+1)}\,\ell$ is proportional
to $|a_n|^2$.

Eq. (\ref{54}) for $l=0$, as a second order differential equation, also has another linearly
independent solution, usually denoted by $W_n$ \cite{arfken}. This solution
is normalizable, but diverging as $v\to \pm 1$. Similar the construction
for $U_n^l$, one can generate the solutions for $l\neq 0$, denoted by
$W_n^l$:
\begin{equation}\label{59}
W_n^l(v)= \sqrt{\frac{2}{\pi}\frac{(n-l)!(n+1)}{(n+l+1)!}} (1-v^2)^{l/2} \frac{\d^l}{\d v^l}W_n(v).
\end{equation}
It is seen that these associated solutions are neither normalizable
nor finite within the interval $[-1,1]$. Further, as the original $W_n$'s are
not in polynomial form, it can be seen that $W_n^l\neq 0$ for $n<l$.
It will be seen later that due to this property these functions can not
appear as coefficients in expansion (\ref{58}) in regions containing
sites with radial sites $n<l$.
The behavior of the two solutions and their
associates best can be obtained by the trigonometric function representation of them
\cite{arfken}, namely:
\begin{align}\label{60}
U_n(x)=\frac{\sin(n+1)\alpha}{\sin\alpha},\\
\label{61}
W_n(x)=\frac{\cos(n+1)\alpha}{\sin\alpha},
\end{align}
with $x=\cos\alpha$. To express the solutions in the exponential form one can
define the linear combinations:
\begin{equation}\label{62}
V_{\pm n}^l:=W_n^l \pm \ir\, U_n^l.
\end{equation}
This exponential representations appear useful to express the wave-functions
in terms of oppositely oscillating radial waves. Also, the behavior of the
above functions at large-$n$ beyond their defining interval $-1\leq v \leq 1$
would come in forms of exponentially decreasing and growing functions of $n$.
As we will see in next section, these functions might
appear as the coefficients $a_n$'s in the expansion (\ref{58}) in the regions with
constant potential; for example in tails of the bound-state
solutions of the radial stepwise potentials. In fact, by the
above form the behaviors of the linear combinations
are summarized at large-$n$ as:
\begin{align}\label{63}
n\to\infty:\left\{
\begin{array}{rl}
 |V_{\pm n}^{l}(x)| \propto \exp(\mp n\, \eta),& x>1,\cr\cr
 |V_{\pm n}^{l}(x)| \propto \exp(\pm n\, \eta),& x <-1,
\end{array}\right.
\end{align}
in which $\cosh^{-1}|x|=\eta>0$.

It is helpful to remind the recurrence relation \cite{ryzhik}
\begin{equation}\label{64}
2\,v\, F_n^l = \alpha_{+n}^l F_{n+1}^l + \alpha_{-n}^l F_{n-1}^l
\end{equation}
with $F$'s are either $U$, $W$, or $V_{\pm n}^l$ types, and
\begin{align}\label{65}
\alpha_{+n}^l &=\sqrt{\frac{(n+l+2)(n-l+1)}{(n+1)(n+2)}},\\
\label{66}
\alpha_{-n}^l &=\sqrt{\frac{(n+l+1)(n-l)}{n(n+1)}}.
\end{align}
We mention
\begin{equation}\label{67}
\alpha_{+n}^l=\alpha_{-(n+1)}^l,~~~~
\alpha_{-n}^l=\alpha_{+(n-1)}^l.
\end{equation}
It is easy to check that the above introduced functions not only satisfy the
above identity in their defining domain $-1\leq v \leq 1$, but also
formally on the whole real axes, $-\infty < v <\infty$.
As the kinetic term is linear in $v$, the above identity comes extremely helpful
to obtain the recurrence relations between the coefficients of the trial expansions.
As an illustration, let us consider the case of a free particle.
By (\ref{45}), the energy eigen-function in momentum space takes the form:
\begin{eqnarray}\label{68}
~_l \langle v , \cos ^{-1}\!\tau , \chi \vert E\rangle=\psi_{E,l}(v)\; Y_{l}^{m}(\cos ^{-1}\!\tau , \chi)
\end{eqnarray}
in which $(\cos ^{-1}\tau , \chi)$ specify the direction of the momentum, and $v$ is
related to the momentum by $v=\cos( \ell\, p/(2\hbar)) $.
As for a free particle momentum commutes with the Hamiltonian, its
energy eigen-function is proportional to $\delta$-function in $v$-space.
Using the representation (\ref{57}), for a free particle with  momentum $\mathbf{p}_0$, with
$v_0=\cos( \ell\, p_0/(2\hbar)) $, we have:
\begin{equation}\label{69}
\psi_{El}(v) =c\, \delta(v-v_0) \propto \sum_{n=l}^\infty U_n^l(v) U_n^l(v_0).
\end{equation}
It would be instructive to check the above result by the use of the expansion (\ref{58}).
By the Hamiltonian of free particle,
\begin{eqnarray}\label{70}
H = \dfrac{4\,\hbar ^{2}}{M \ell ^{2}}({1} - { v }),
\end{eqnarray}
and the identity (\ref{64}), the equation $H\psi_E=E\psi_E$ would lead to the
recurrence relation for the coefficients in the expansion (\ref{58}):
\begin{eqnarray}\label{71}
2 \left( 1 - \dfrac{M \ell ^{2} E}{4\, \hbar ^2}\right) a_n
= \alpha _{-n}^{l} a_{n-1} + \alpha _{+n}^{l}a_{n+1},
\end{eqnarray}
with the boundary condition $a_{l-1}=0$. Defining
\begin{eqnarray}\label{72}
 v_0 := 1 - \dfrac{M \ell ^{2} E}{4 \,\hbar ^2},
\end{eqnarray}
and a fresh use of the identity (\ref{64}),
we find $a_n \propto U_{n}^{l}( v _0)$, as
confirmation of the result (\ref{69}).
In an alternative way, one may choose the linear combination
\begin{equation}\label{73}
a_n=C^+\, V_{+n}^l(v_0) +C^-\,V_{-n}^l(v_0), ~~~~n=l,l+1,\cdots,
\end{equation}
for which by the condition $a_{l-1}=0$, we find
$C^+=-C^-$, leading to the previous result.
By the condition $-1\leq v_0 \leq 1$ for detectable particles, we find
\begin{equation}\label{74}
0 \leq E \leq \dfrac{8\hbar ^{2}}{M \ell ^{2}},
\end{equation}
expressing that the energy of a free particle with its momentum taking values on a
compact space would be bounded from above.

\section{Radial stepwise potential}
By the quantized radial distance as $r_n=\sqrt{\frac{n}{2}(\frac{n}{2}+1)}\,\ell $, the radial stepwise potentials may be defined by
\begin{eqnarray}\label{75}
V(r_n) = \left\{\begin{array}{rcl}
\pm V_{0}  ,& n \leq n_{0}& \mathrm{(region~I)}\\~\\
0 , &n > n_{0}& \mathrm{(region~II)}
\end{array}\right.
\end{eqnarray}
in which $+V_0$ and $-V_0$ correspond to the radial barrier and the radial square well
potentials, respectively. As mentioned for the free particle, the total energy eigen-function
in momentum space has the form (\ref{68}), for which the dependence on
$v$ can be expanded as (\ref{58}). For the regions~I and II of the potential,
the recurrence relations for the coefficients are found as below:
\begin{align}\label{76}
n\leq n_0:~~&2 \,v_\mathrm{I}\,
a_n =  \alpha _{-n}^{l} a_{n-1} + \alpha _{+n}^{l}a_{n+1} \\
\label{77}
n> n_0:~~&2 \,v_\mathrm{II}\,
a_n =  \alpha _{-n}^{l} a_{n-1} + \alpha _{+n}^{l}a_{n+1} ,
\end{align}
in which
\begin{align}\label{78}
v_{\mathrm{I}}& :=1 - \dfrac{M \ell ^2}{4 \,\hbar ^{2}}(E \mp  V_0 ),\\
\label{79}
v_{\mathrm{II}} &:= 1 -\dfrac{M \ell^{2}E}{4\,\hbar ^{2}},
\end{align}
accompanied by the boundary condition $a_{l-1}=0$.
By the condition $n\geq l$ mentioned in the previous section,
the more interesting cases happen when $l<n_0$, for which
the recurrence relation in region~I has nonzero solution.
By the properties of $U_n^l$'s mentioned before and
the boundary condition, for the region~I the acceptable solution
comes in the form
\begin{align}\label{80}
a_n= C_\mathrm{I} ~ U_{n}^{l}( v_{\mathrm{I}}),~~~~~~n=l,\cdots, n_0+1
\end{align}
For the region~II, based on the behavior of the eigen-function for
$n\to \infty$, either $V_{\pm n}^l$ with the argument inside
the interval $[-1,1]$, or exponentially decreasing $V_{+n}^l$ type
outside the interval by (\ref{63}), are acceptable solutions.

Based on the condition $|v|\leq 1$ for directly detectable particles,
two different situation should be studied separately, which are:
1) $8\,\hbar^2/(M\,\ell^2)>V_0$, and
2)  $8\,\hbar^2/(M\,\ell^2)<V_0$.

\subsection{Case with $8\,\hbar^2/(M\,\ell^2)>V_0$}
Here we consider the barrier and square well cases separately. \\

\underline{\textbf{Barrier case:}}\\
In this case three domains for $E$ are recognized, for each one
the corresponding $v_\mathrm{I\,\&\,II}$ are mentioned:
\begin{align}\label{81}
0\leq E \leq V_0:&
~~  v_\mathrm{I}\geq 1 ~~~\mathrm{and}~~ |v_{\mathrm{II}}| \leq 1\cr
V_0\leq E \leq \frac{8\,\hbar^2}{M\,\ell^2} :&
~~  |v_\mathrm{I}|\leq 1 ~~\mathrm{and}~~ |v_{\mathrm{II}}| \leq 1\cr
\frac{8\,\hbar^2}{M\,\ell^2}\leq E \leq\frac{8\,\hbar^2}{M\,\ell^2}+V_0:&
~~  |v_\mathrm{I}|\leq 1 ~~\mathrm{and}~~ v_{\mathrm{II}} \leq -1
\end{align}
Out of three domains mentioned in above we have $|v_\mathrm{I\,\&\,II}|>1$,
which are not acceptable for a particle detectable in either region~I or II. For
the first two domains in above, $v_\mathrm{II}$ takes the values for which
in region~II the particle can make oppositely oscillating waves. So
\begin{align}\label{82}
0\leq E \leq\frac{8\,\hbar^2}{M\,\ell^2}:~~
 a_n=C^+_\mathrm{II} \, V_{+n}^l(v_{\mathrm{II}}) +C^-_\mathrm{II}\,
 V_{-n}^l(v_{\mathrm{II}}), ~~~~ n=n_0,\cdots,\infty
\end{align}
The continuity condition between two regions at $n_0$ and $n_0+1$ would give the relations between three
pre-factors $C_\mathrm{I}$ and $C^\mathrm{\pm}_\mathrm{II}$,
and no condition on energy would be required. As the consequence,
in the domains the energy spectrum is continuous. We mention that
this domain of energy reaches the upper bound for the energy of a free particle obtained
in the previous section.

For the third domain for energy, however, the particle can not have a propagating nature in the region~II, and so the wave-function should vanish exponentially as $n\to\infty$. So, 
by $v_\mathrm{II}<-1$ in third domain and (\ref{63}), we have
\begin{align}\label{83}
\frac{8\,\hbar^2}{M\,\ell^2}\leq E \leq\frac{8\,\hbar^2}{M\,\ell^2}+V_0:~~
 a_n=C^-_\mathrm{II} \, V_{-n}^l(v_{\mathrm{II}}) , ~~~~ n=n_0,\cdots,\infty
\end{align}
In this case the two continuity equations, namely
\begin{align}\label{84}
 C_{\mathrm{I}}~U_{n_{0}}^{l}( v_{\mathrm{I}}) &= C^-_{\mathrm{II}}~
V_{-n_{0}}^{l}( v_{\mathrm{II}}) \\
\label{85}
  C_{\mathrm{I}}~U_{n_{0}+1}^{l}( v_{\mathrm{I}}) &= 
C^-_{\mathrm{II}}~V_{-(n_{0}+1)}^{l}( v_{\mathrm{II}}) .
\end{align}
 are sufficient to fix the relation between two
pre-factors $C_\mathrm{I}$ and $C^+_\mathrm{II}$, provided that the determinant
of the equations would vanish, leading to
\begin{equation}\label{86}
U_{n_{0}}^{l}( v_{\mathrm{I}}) \,V_{-(n_{0}+1)}^{l}( v_{\mathrm{II}}) -
V_{-n_{0}}^{l}( v_{\mathrm{II}})\, U_{n_{0}+1}^{l}( v_{\mathrm{I}}) = 0.
\end{equation}
The last expression is in fact the quantization condition, by which a discrete set
of the energies in the third interval is obtained.
As for these kinds of solutions
$V_{-n}^l(v_{\mathrm{II}})\to 0$ by $n\to\infty$, these states with discrete
energies are bound-ones. The surprising feature of
these bound-states is that they are obtained with an initially supposed repulsive potential
of a barrier. It is in fact the result of the bound on the momentum.
In particular, although it is expected that outside the repulsive region~I the momentum
would grow, but due to the bound, outside the repulsive region would appear
forbidden for particle to propagate. \\

\underline{\textbf{Square well case:}}\\
Also in this case three domains for $E$ are recognized:
\begin{align}\label{87}
-V_0\leq E \leq 0:&
~~  |v_{\mathrm{I}}| \leq 1 ~~~\mathrm{and}~~ v_\mathrm{II}\geq 1\cr
0\leq E \leq -V_0+\frac{8\,\hbar^2}{M\,\ell^2} :&
~~  |v_\mathrm{I}|\leq 1 ~~\mathrm{and}~~ |v_{\mathrm{II}}| \leq 1\cr
-V_0+\frac{8\,\hbar^2}{M\,\ell^2}\leq E \leq\frac{8\,\hbar^2}{M\,\ell^2}:&
~~  v_\mathrm{I}\leq -1 ~~\mathrm{and}~~ |v_{\mathrm{II}}| \leq 1
\end{align}
For the first interval, the acceptable solutions in region~II are of the form of
(\ref{83}), but with $V_{-n}$ replaced by $V_{+n}$. So the quantization 
condition comes to the form 
\begin{equation}\label{88}
U_{n_{0}}^{l}( v_{\mathrm{I}}) \,V_{+(n_{0}+1)}^{l}( v_{\mathrm{II}}) -
V_{+n_{0}}^{l}( v_{\mathrm{II}})\, U_{n_{0}+1}^{l}( v_{\mathrm{I}}) = 0.
\end{equation}
In this case the energy spectrum is discrete, and the eigen-functions, as expected, 
are bound-states.

For the second and third energy intervals the acceptable solutions are of the
form of (\ref{82}), and we encounter with asymptotically free states.
The surprising feature in this case is with the third interval, for which,
although the energy is higher, but due to the bound on the maximum
momentum, it is forbidden for the particle to propagate in the attractive
region~I.

\subsection{Case with $8\,\hbar^2/(M\,\ell^2)<V_0$}
Here also we consider the barrier and square well cases separately. \\

\underline{\textbf{Barrier case:}}\\
In this case two domains for $E$ are recognized, for each one
the corresponding $v_\mathrm{I\,\&\,II}$ are mentioned:
\begin{align}\label{89}
0\leq E \leq \frac{8\,\hbar^2}{M\,\ell^2} :&
~~  v_\mathrm{I}> 1 ~~\mathrm{and}~~ |v_{\mathrm{II}}| \leq 1\cr
V_0\leq E \leq\frac{8\,\hbar^2}{M\,\ell^2}+V_0:&
~~  |v_\mathrm{I}|\leq 1 ~~\mathrm{and}~~ v_{\mathrm{II}} \leq -1
\end{align}
By the same reasonings of the previous part, in the first interval the energy spectrum is
continuous and in the second one is discrete (by condition (\ref{86})). 
Also, again due to the bound
on momentum, in the second interval region~II is forbidden for
particle to propagate, and so states are surprisingly bound-ones.
Also in this case there is a gap in the interval
$[\frac{8\,\hbar^2}{M\,\ell^2},V_0]$
between the continuous (lower) and the discrete (higher) 
parts of the spectrum. \\

\underline{\textbf{Square well case:}}\\
Also in this case two domains for $E$ are recognized:
\begin{align}\label{90}
-V_0\leq E \leq -V_0+\frac{8\,\hbar^2}{M\,\ell^2} :&
~~  |v_\mathrm{I}|\leq 1 ~~\mathrm{and}~~ v_{\mathrm{II}}>1\cr
0\leq E \leq\frac{8\,\hbar^2}{M\,\ell^2}:&
~~  v_\mathrm{I}\leq -1 ~~\mathrm{and}~~ |v_{\mathrm{II}}| \leq 1
\end{align}
It is easy to see that in the first interval the energy spectrum is
discrete (by condition (\ref{88})), 
and in the second one is continuous. 
Also, due to the bound
on momentum, in the second interval the particle can
not propagate in the region~I, although this region is
attractive. Also in this case there is a gap in the interval
$[-V_0+\frac{8\,\hbar^2}{M\,\ell^2},0]$
between the discrete (lower) and the continuous (higher) 
parts of the spectrum.

\subsection{Numerical samples}
The spectrum of energy for stepwise potentials obtained in
the previous subsections can be checked in both discrete and continuous
parts by approximate methods; for example by the basis $\{U_n^l\}$ in
the Rayleigh-Ritz perturbation method \cite{merzbacher}.
Here for the discrete part of spectrum we give the samples of
the numerical solutions for (\ref{86}) or (\ref{88}), for both barrier
and square well potentials, presented in Tables 1~\&~2, respectively.

\begin{table}[t]{\scriptsize
\begin{center}
\begin{tabular}{c|ccccccc}
  & $E_{0}$  & $E_{1}$  & $E_{2}$  & $E_{3}$  & $E_{4}$   & $E_{5}$ & $E_{6}$ \\
\hline
$l=0$ & 13.851 & 13.425  & 12.785 & 12.025 & 11.257  & 10.599 & 10.156 \\
$l=1$ & --& 13.696 & 13.144 & 12.419 & 11.628 & 10.889 & 10.318 \\
$l=2$ & --& -- & 13.497  & 12.823 & 12.025 & 11.218 & 10.525 \\
$l=3$ & --& -- & -- & 13.250 & 12.455 & 11.592 & 10.781  \\
\end{tabular}
\caption{\small The numerical values for the bound-states energy values
by the quantization condition (\ref{86}) for the barrier potential.
The presented values are for $4 \hbar ^{2}/(M \ell ^{2}) =2$, $n_{0}=6$, $V_{0}=10$. }
\end{center}
}\end{table}

\begin{table}[t]{\scriptsize
\begin{center}
\begin{tabular}{c|ccccccc}
  & $E_{0}$  & $E_{1}$  & $E_{2}$  & $E_{3}$  & $E_{4}$   & $E_{5}$ & $E_{6}$ \\
\hline
$l=0$ & -9.8510 & -9.4258  & -8.7858 & -8.0251 & -7.2571  & -6.5995 & -6.1564 \\
$l=1$ & --& -9.6960 & -9.1448 & -8.4199 & -7.6283 & -6.8896 & -6.3185 \\
$l=2$ & --& -- & -9.4972  & -8.8239  & -8.0251 & -7.2183 & -6.5254 \\
$l=3$ & --& -- & -- & -9.2502 & -8.4558 & -7.5920 & -6.7810  \\
\end{tabular}
\caption{\small The numerical values for the bound-states energy values
by the quantization condition (\ref{88}) for the square-well potential.
The presented values are for $4 \hbar ^{2}/(M \ell ^{2}) =2$, $n_{0}=6$, $V_{0}=10$. }
\end{center}
}\end{table}

\section{General potential}
Here we consider the case with a general rotationally invariant potential.
As the result, it is shown that in general the eigenvalue problem would
end to solve a 3-term recurrence relation, or equivalently to find
the stable points of a corresponding continued fraction. Using
\begin{equation}\label{91}
(\mathbf{x} \cdot \mathbf{x})\, \vert n \rangle_l = \dfrac{\ell ^{2}}{4} n(n+2)\, \vert  n \rangle_l
\end{equation}
we have
\begin{equation}\label{92}
{}_l\langle n\vert\, V(\sqrt{\mathbf{x \cdot x}})\,\vert\psi\rangle =  V\left( \sqrt{\dfrac{n}{2}\left(\dfrac{n}{2}+1\right)}\,\ell \right) \psi_l(n) .
\end{equation}
By using the identity (\ref{64}), the equation $H\psi_E=E\psi_E$ would lead to the
recurrence relation for the coefficients of expansion (\ref{58}):
\begin{eqnarray}\label{93}
2\left(1 - \dfrac{M \ell ^{2}}{4\, \hbar ^{2}}\left(E- V_n\right)\right) a_n =
\alpha^{l}_{-n} a_{n-1}+ \alpha^{l}_{+n} a_{n+1}
\end{eqnarray}
in which
\begin{equation}\label{94}
V_n:=V\left( \sqrt{\dfrac{n}{2}\left(\dfrac{n}{2}+1\right)}\,\ell \right),
\end{equation}
accompanied by the boundary  condition $a_{l-1}=0$. 
The treatment of these kinds of recurrence relations is 
rather standard \cite{3term,jeffreys}. Defining
\begin{equation}\label{95}
 N_{n}:=\dfrac{a_{n+1}}{a_n}
\end{equation}
the above 3-term recurrence relation can be transformed to 
\begin{equation}\label{96}
N_{n-1} = \dfrac{\alpha^{l}_{-n}}{2\left(1 - \dfrac{M \ell ^{2}}{4\, \hbar ^{2}}
\left(E - V_n\right)\right) -\alpha^{l}_{+n} N_{n}}
\end{equation}
with $n=l,l+1,\cdots$. We mention, following the condition $a_{l-1}=0$, $N_{l-1}\to\infty$.
As consequence, by setting $n=l$ in above the denominator should vanish, leading to
\begin{equation}\label{97}
N_l=\sqrt{2(l+2)}
\left(1 - \dfrac{M \ell ^{2}}{4\, \hbar ^{2}} \left(E - V_l\right)\right).
\end{equation}
On the other hand, one can express $N_l$ by means of
continued fractions, namely
\begin{equation}\label{98}
N_{l} = \dfrac{\alpha^{l}_{+l}}{\beta _{l+1}-
\dfrac{\left(\alpha^{l}_{+(l+1)}\right)^{2}}{\beta _{l+2}-
\dfrac{\left(\alpha _{+(l+2)}\right)^{2}}{\beta _{l+3} -\cdots}}}
\end{equation}
in which we have used the relation $\alpha_{-(n+1)}^l=\alpha_{+n}^l$, and defining
\begin{eqnarray}\label{99}
\beta_{n} := 2\left(1 - \dfrac{M \ell ^{2}}{4 \,\hbar ^{2}}\left(E - V_n\right)\right).
\end{eqnarray}
In practice, firstly one should determine the limiting value of
$N_n$ for $n\to\infty$. Then by equating the two values for $N_l$ by (\ref{97})
and a truncated form of (\ref{98}) by the limiting value, one can get
an equation by which the energy eigenvalues could be evaluated.
The accuracy as well as the number of obtained eigenvalues
would be determined by the level of truncation of the continued fraction (\ref{98}).
Hence, the desired accuracy could be reached by sufficiently large level of
truncation \cite{jeffreys}, say $n_\infty$, by which (\ref{98}) takes the form
\begin{equation}\label{100}
N_l=  \dfrac{\alpha^{l}_{+l}}{\beta _{l+1}\,-}
\dfrac{\left(\alpha^{l}_{+(l+1)}\right)^{2}}{\beta _{l+2}\, -}
\cdots
\dfrac{\left(\alpha^{l}_{+(l+n_\infty-1)}\right)^{2}}{\beta _{l+n_\infty} -\alpha^l_{+(l+n_\infty)} N_{n_\infty}} .
\end{equation}
Once the energy eigenvalues are determined by the desired accuracy,
the eigen-functions can be constructed by solving the recurrence relations
(\ref{93}) for $a_n$'s, accompanied by appropriate boundary and
normalization conditions.

In the following we apply this method to the cases with the harmonic oscillator and the coulomb potentials.

\subsection{Harmonic oscillator}
The harmonic oscillator potential is taken as
$\dfrac{1}{2}M \omega^2 (\mathbf{x \cdot x})$, by which
we have
\begin{equation}\label{101}
N_l=\sqrt{2(l+2)}
\left(1 - \dfrac{M \ell ^{2}}{4\, \hbar ^{2}}
\left(E - \dfrac{1}{8}M \omega^2 \ell^{2} l(l+2)\right)\right).
\end{equation}
At large $n$ we assume $N_n\propto c\, n^\gamma$,
by which after inserting in (\ref{96}), we find
\begin{equation}\label{102}
\gamma=-2,~~~c=\frac{16\,\hbar^2}{M^2\ell^4 \omega^2},
\end{equation}
leading to $N_n\to 0$ as $n\to\infty$. So truncation at
level $n_\infty$ would be hold by $N_{n_\infty}\approx 0$ in (\ref{100}).
Equating (\ref{101}) and the truncated form of (\ref{100}) leads
to the equation for eigenvalues.

In Table~3 samples of the numerical solutions by the method as
the energy eigenvalues are given. All the given numbers can be checked
also by the approximation methods, for example the Rayleigh-Ritz method.

\begin{table}[t]{\scriptsize
\begin{center}
\begin{tabular}{c|ccccccc}
  & $E_{0}$  & $E_{1}$  & $E_{2}$  & $E_{3}$  & $E_{4}$   & $E_{5}$ & $E_{6}$  \\
\hline
$l=0$ & 1.6867 & 5.1130 &  10.057 & 17.031 & 26.020 & 37.014 & 50.010 \\
$l=1$ & - &4.8683 &10.012 & 17.018 & 26.015 & 37.011 & 50.009  \\
$l=2$ & - & -  & 9.9288 & 16.993  & 26.005 & 37.006 & 50.006 \\
$l=3$ & - & - & - & 16.955 & 25.989 & 36.999 & 50.002
\end{tabular}
\caption{\small The energy eigenvalues for the harmonic oscillator, setting $4 \hbar ^{2}/(M \ell ^{2}) =2$, $\hbar\omega=2 $, level of truncation: 14.}
\end{center}
}\end{table}

\subsection{Coulomb potential}
The Coulomb potential is taken as
$ V (\sqrt{\mathbf{x \cdot x}}) = \dfrac{- e^{2}}{\sqrt{\mathbf{x \cdot x}}}$, 
by which we have
\begin{equation}\label{103}
N_l=\sqrt{2(l+2)}
\left(1 - \dfrac{M \ell ^{2}}{4\, \hbar ^{2}}
\left(E + \frac{2\,e^2}{\ell \sqrt{ l(l+2)}}\right)\right).
\end{equation}
In this case we have both the bound-states for $E<0$, as well as
the asymptotically free states with $E>0$.
Here we consider only the case $E<0$.
At large $n$ we assume $N_n\propto c\, n^\gamma$,
by which after inserting in (\ref{96}), it would be found
\begin{equation}\label{104}
\gamma=0,~~~c= 1- \dfrac{M\ell^{2}}{4\,\hbar^{2}}
\left(E \pm \sqrt{E ^2 -\dfrac{8\,\hbar^{2}}{M\ell^{2}}E}\right)=:c_\pm.
\end{equation}
As far as the aim is to find the energy eigenvalues,
both $c_\pm$ can be used as the limiting value
in (\ref{100}). So, equating (\ref{103}) and truncated form
of (\ref{100}) by $N_{n_\infty}\approx c_\pm$,
would yield the required equation for the discrete eigenvalues
of bound-states. However, to obtain the coefficients in the expansion
the situation is different. It is easy to check
\begin{equation}\label{105}
c_+\,c_-=1,~~~~ c_+>c_-,
\end{equation}
by which we have $c_+>1$ and $c_-<1$. As $N_n$ is giving the
ratio $a_{n+1}/a_n$, to have normalizable bound-states of the form
(\ref{58}) only $c_-$ can be accepted to solve the recurrence relation (\ref{93}).

In Table~4 samples of the numerical solutions by the method as
the energy eigenvalues are given. Also in this case the given numbers can be checked
by the approximation methods, for example the Rayleigh-Ritz method.

\begin{table}[t]{\scriptsize
\begin{center}
\begin{tabular}{c|ccccccc}
  & $E_{0}$  & $E_{1}$  & $E_{2}$  & $E_{3}$  & $E_{4}$   & $E_{5}$ & $E_{6}$  \\
\hline
$l=0$ & $-\infty$ & -16.614 &  -9.5004 & -6.5087 & -4.8372 & -3.7717 & -3.0371 \\
$l=1$ & - & -16.568 & -9.4921 & -6.5056 & -4.8357 & -3.7708 & -3.0366  \\
$l=2$ & - & -  & -9.4766 & -6.4995  & -4.8326 & -3.7690 & -3.0355 \\
$l=3$ & - & - & - & -6.4908 & -4.8283 & -3.7666 & -3.0340
\end{tabular}
\caption{\small The energy eigenvalues for the Coulomb potential, setting $4 \hbar ^{2}/(M \ell ^{2}) =2$, $e^2 / \ell = 16 $, level of truncation: 14. }
\end{center}
}\end{table}

\vskip 0.5cm
\noindent\textbf{Acknowledgement}:  This work is
supported by the Research Council of Alzahra University.
The authors are grateful to M.~Khorrami for helpful discussions.

\appendix
\section{Samples of $U_n^l$'s and their plots}
Here some explicit expressions of the
associated Chebyshev polynomials $U_{n}^{l} $'s, and their
plots are presented. \\

\begin{tabular}{l}
Associated Chebyshev functions $U_{n}^{1}(x) $\\
\hline\cr
$U_{1}^{1}(x) = 2\sqrt{\frac{2}{3\pi}}\sqrt{1-x^2}$ \\
$U_{2}^{1}(x) =\frac{1}{\sqrt{\pi}} 4x\sqrt{1 - x^2} $ \\
$U_{3}^{1}(x) =  \sqrt{\frac{2}{15\pi}}\sqrt{1-x^2}(24x^{2} - 4)$ \\
$U_{4}^{1}(x) = \frac{1}{2\sqrt{3\pi}}\sqrt{1 - x^2}( 64x^{3} - 24x)$ \\
$U_{5}^{1}(x) =  \sqrt{\frac{2}{35\pi}}\sqrt{1-x^2}(160x^{4} - 96 x^{2} + 6)$ \\
$U_{6}^{1}(x) = \frac{1}{2\sqrt{6\pi}}\sqrt{1 - x^2}(384 x^{5} - 320 x^{3} + 48 x)$\\
\end{tabular} \\ 

\begin{figure}[h]
\label{fig2}
\centerline{\includegraphics[scale=0.35]{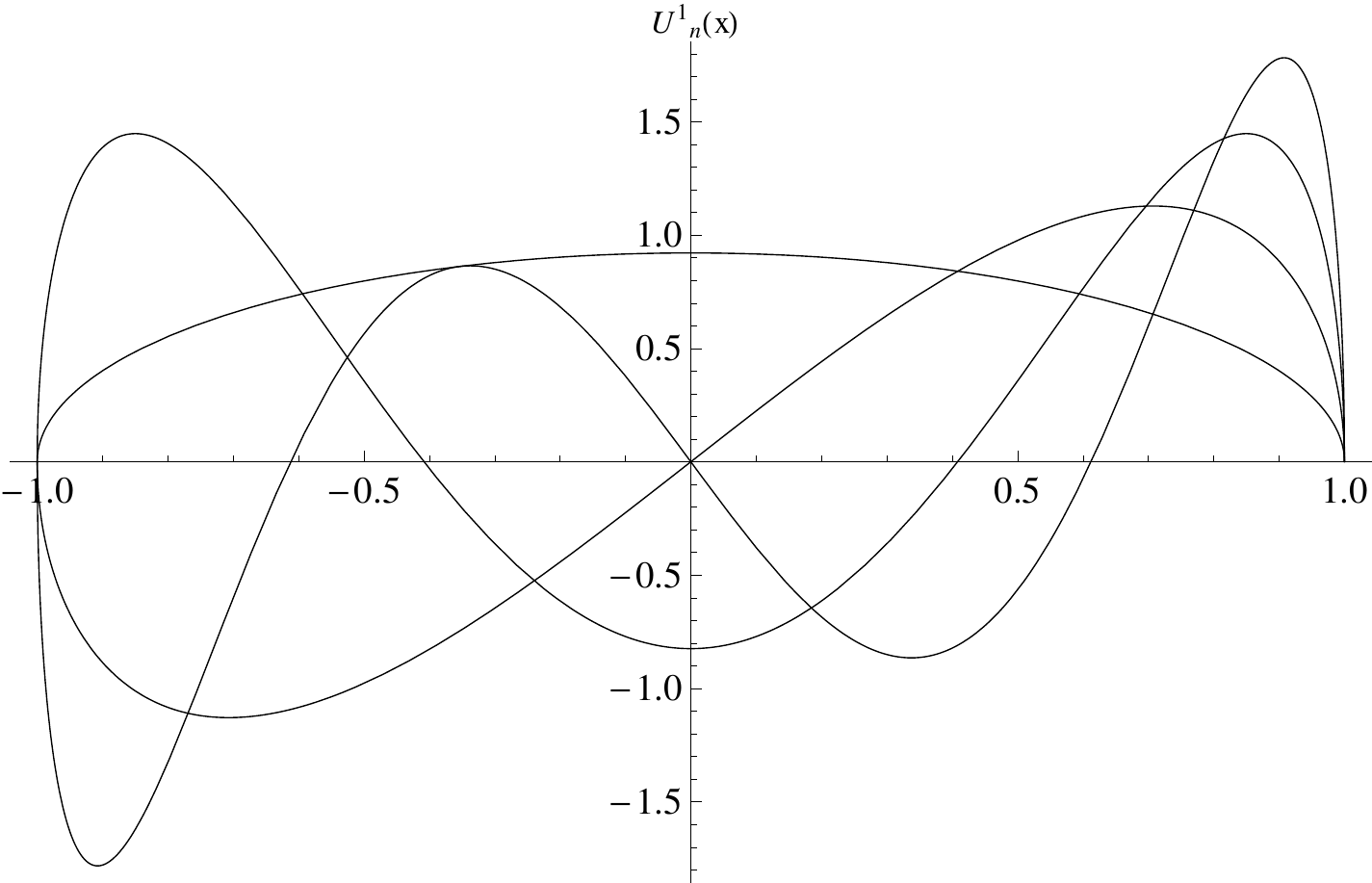}}
\caption{Plots $U_{n}^{1}(x)$, $n=1,2,3,4$.  }
\end{figure}

\begin{tabular}{l}
Associated Chebyshev functions $U_{n}^{2}(x) $\\ 
\hline\cr
$U_{2}^{2}(x) = \frac{4}{\sqrt{5\pi}}(1 - x^2)$ \\
$U_{3}^{2}(x) =8\sqrt{\frac{2}{5\pi}}x(1 - x^2) $ \\
$U_{4}^{2}(x) = \frac{1}{6\sqrt{7\pi}}(1- x^2)(192 x^2 -24)$ \\
$U_{5}^{2}(x) = \frac{1}{4\sqrt{35\pi}}(1 - x^2)( 640x^{3} - 192 x)$ \\
$U_{6}^{2}(x) = \frac{1}{6\sqrt{30\pi}} (1 - x^2)(1920 x^4 - 960 x^2 + 48) $ \\
$U_{7}^{2}(x) = \frac{1}{3\sqrt{210\pi}}(1 - x^2)(5376 x^{5} - 3840 x^{3} + 480 x)$\\
\end{tabular}

\begin{figure}[h]
\label{fig3}
\centerline{\includegraphics[scale=0.35]{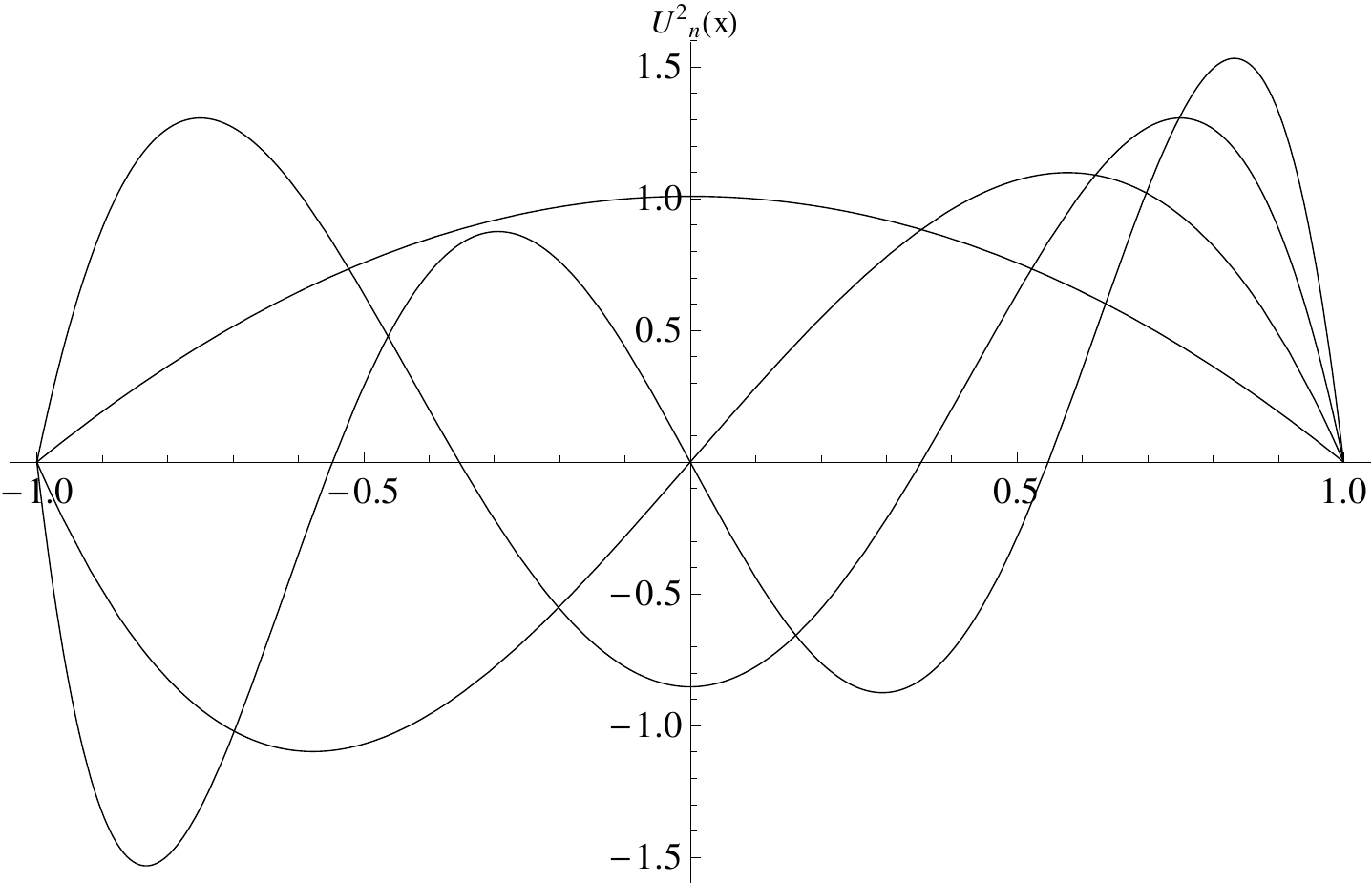}}
\caption{Plots of $U_{n}^{2}(x)$, $n=1,2,3,4$.  }
\end{figure}

\end{document}